\documentclass[preprintnumbers,amsmath,amssymb,prd,twocolumn,nofootinbib]{revtex4}
\usepackage{amsmath,amssymb,mathrsfs}
\renewcommand{\vec}[1]{\boldsymbol{\mathrm{#1}}}
\usepackage{graphicx}

\begin{document}
\title{Rotational Velocity Curves in the Milky Way as a Test of Modified Gravity}

\author{J. W. Moffat$^{\dag\star}$ and V. T. Toth$^{\dag}$\\~\\
$^{\dag}$Perimeter Institute for Theoretical Physics,\\
Waterloo ON N2L 2Y5, Canada\\
$^\star$Department of Physics and Astronomy, University of Waterloo,\\
Waterloo ON N2L 3G1, Canada}



\begin{abstract}
Galaxy rotation curves determined observationally out to a radius well beyond the galaxy cores can provide a critical test of modified gravity models without dark matter. The predicted rotational velocity curve obtained from Scalar-Vector-Tensor Gravity (STVG or MOG) is in excellent agreement with data for the Milky Way without a dark matter halo, with a mass of $5\times 10^{10}\,M_{\odot}$. The velocity rotation curve predicted by modified Newtonian Dynamics (MOND) does not agree with the data.
\end{abstract}

\maketitle

\section{Introduction}

The Scalar-Tensor-Vector Gravity (STVG) or modified gravitational (MOG) theory~\cite{Moffat2006a} has successfully explained the rotation curves of galaxies~\cite{Brownstein2007,MoffatRahvar2013} and the dynamics of galactic clusters~\cite{MoffatRahvar2014}, as well as describing the growth of structure, the matter power spectrum and the cosmic microwave background (CMB) acoustical power spectrum~\cite{Moffat2014a}. The modified Newtonian acceleration law obtained in the theory in the weak field, non-relativistic approximation to the field equations, reduces to Newtonian gravity for the solar system describing solar system experiments in agreement with general relativity (GR).

Galaxies with observational data for rotational velocity curves reaching out well beyond the galaxy cores can provide a critical test of modified gravity theories without dark matter as well as dark matter models. Observed rotational velocity data has been obtained for the Milky Way that extends as far as $200\,{\rm kpc}$ from the core~\cite{Bhattacharjee2014}. The rotational velocity predicted by STVG asymptotically becomes Kepler-Newtonian with an increased value for the gravitational constant $G=G_{\infty}$. On the other hand, the MOND prediction for the rotational velocity curve becomes asymptotically constant for a value of the critical acceleration $a_0\sim 10^{-8}\,{\rm cm\, s^{-2}}$. The Milky Way data can be fitted well by STVG but it excludes MOND as a viable model that can fit galaxy rotational velocity data.

\section{MOG and MOND Acceleration laws}

The effective potential for an extended distribution of matter in MOG in the weak field approximation is given by~\cite{MoffatRahvar2013}:
\begin{align}
\Phi_{\rm eff}(\vec x) = - G_N\int\frac{\rho(\vec x')}{|\vec x-\vec x'|}\left[1+\alpha-\alpha e^{-\mu|\vec x-\vec x'|}\right]d^3\vec{x}',
\label{eq:MOG1}
\end{align}
where $G_N$ is Newton's gravitational constant, while $\mu$ and $G$ are scalar fields, the latter defined by $\alpha = (G - G_N)/G_N$ using the notation given by Moffat and Toth~\cite{Moffat2007e}.

The MOG acceleration of a test particle can be obtained from the gradient of the potential, $\vec{a} = -\vec{\nabla}\Phi_{\rm eff}$, yielding the
result:
\begin{align}
\vec{a}(\vec x) =& - G_N\int\frac{\rho(\vec x')(\vec{x}-\vec{x'})}{|\vec x-\vec x'|^3}\nonumber\\
&{}\times\left[1+\alpha -\alpha e^{-\mu|\vec x-\vec x'|}(1+\mu|\vec x-\vec x'|) \right]d^3\vec{x}'. \label{acceleration}
\end{align}
We note that MOG has a remarkable predictive power for rotation curves of galaxies~\cite{MoffatRahvar2013} and for the dynamics of galaxy clusters~\cite{MoffatRahvar2014}. The values of $\alpha$ and $\mu$ can be obtained as functions of the mass of a point gravitational source \cite{Moffat2007e}. Alternatively, Moffat and Rahvar used the ``universal'' values of $\alpha=8.89$ and $\mu=0.04\,{\rm kpc}^{-1}$ for given values of mass-to-light ratio $\Upsilon=M/L$ to fit the rotation curves of several galaxies. In either case, the values of $\alpha$ and $\mu$ allow for parameter-free fits to rotation curves and to the isothermal mass profiles of galaxy clusters without dark matter. The general Tully-Fisher scaling relation for galaxies is well predicted by MOG.

To determine rotational velocity curves well beyond the galaxy core, we can use the MOG point particle acceleration law~\cite{Moffat2006a,Moffat2007e}:
\begin{equation}
\label{MOGacceleration}
a(r)=\frac{G_NM}{r^2}\Big[1+\alpha-\alpha(1+\mu r)e^{-\mu r}\Big],
\end{equation}
where the acceleration is in the direction of the source.

For rotational velocity curves well beyond the galaxy core, we can use the MOND non-relativistic acceleration law given by~\cite{Milgrom1983}:
\begin{equation}
a(r)=\frac{a_N(r)}{\mu(a/a_0)},
\end{equation}
where $a_N(r)$ is the Newtonian acceleration, $a_N(r)=G_NM/r^2$ and $\mu(x)$ is the interpolation function (not to be confused with the MOG $\mu$ parameter in Eq.~(\ref{eq:MOG1})) whose asymptotic values are $\mu(x)=1$ when $a\gg a_0$ and $\mu(x)=a/a_0$ when $a\ll a_0$. The quantity $a_0$ is the critical acceleration below which the Newtonian gravity is no longer valid. Studies have found that $a_0\sim 1.21\times 10^{-8}\, {\rm cm\,s^{-2}}$~\cite{Begeman1991,Brada1995,Sanders1996}. The interpolation function is usually given by
\begin{equation}
\mu(x)=\frac{x}{\sqrt{1+x^2}}.\label{MOND1}
\end{equation}
It is clear that there is a large family of $\mu(x)$ functions which are compatible with the asymptotic behaviors. An alternative example is~\cite{Famaey2005}:
\begin{equation}
\mu(x)=\frac{x}{1+x}.\label{MOND2}
\end{equation}

\section{Fits to Rotational Velocity Curve for the Galaxy}

In Ref. \cite{Bhattacharjee2014}, high quality observations of the rotation curve of the Milky Way have been obtained to a radial distance of 200~kpc using a heterogeneous set of objects including globular clusters.

This rotation curve offers an opportunity to test various gravitational theories, including MOG and MOND. Furthermore, as we investigate rotation curves at significantly larger radii than the central region of the galaxy where nearly all baryonic mass resides, it is justifiable to use a point source approximation for any physical model that does not involve an extended dark matter halo.

Accordingly, we obtained the rotational velocity curve by using the circular velocity formula:
\begin{equation}
v_c(r)=\sqrt{a(r)r}.
\end{equation}

We investigated five scenarios: MOG using two different sets of the parameters $\alpha$ and $\mu$ (one set obtained using the solution in \cite{Moffat2007e}, another set obtained by fitting a series of disk galaxies \cite{MoffatRahvar2013}); MOND using the two different interpolation functions given in Eqs.~(\ref{MOND1}) and (\ref{MOND2}); and for comparison, a point-source Newtonian fit using an appropriately adjusted (but astrophysically unrealistic) source mass.

In addition, we also included rotation curve estimates using a fitted dark matter halo \cite{Sofue2012}. For an extended halo, a point source approximation is clearly inappropriate. Instead, we used the six-parameter formulation given in \cite{Sofue2012}, which expresses the circular velocity in terms of contributions from the bulge, disk, and halo:
\begin{align}
v_c(r)^2=&\frac{GM_b}{a_b}\frac{4}{\eta}\mathscr{B}(r/a_b)^2+\frac{GM_d}{a_d}{\cal D}(r/a_d)^2\nonumber\\
&{}+\frac{4\pi G\rho_0h^3[\log(1+r/h)+r/(r+h)]}{r},\label{eq:Sof1}
\end{align}
where the helper functions $\mathscr{B}(x)$ and ${\cal D}(x)$ are given by\footnote{The factor of $4/\eta$ in Eq.~(\ref{eq:Sof1}) appears to have been inadvertently omitted from \cite{Sofue2012}; furthermore, in the same reference, Eq.~(\ref{eq:Sof2}) appears to have the erroneous $t^2-1$ instead of $t^2-y^2$ under the square root in the denominator.}
\begin{align}
\mathscr{B}(r)^2&=r^{-2}\int_0^ry^2\int_y^\infty\frac{\displaystyle\frac{d}{dt}e^{-\kappa(t^{1/4}-1)}}{\sqrt{t^2-y^2}}dtdy,\label{eq:Sof2}\\
\mathscr{D}(x)^2&=\textstyle\frac{1}{2}x\Big[I_0\left(\textstyle\frac{1}{2}x\right)K_0\left(\textstyle\frac{1}{2}x\right)-I_1\left(\textstyle\frac{1}{2}x\right)K_1\left(\textstyle\frac{1}{2}x\right)\Big],
\end{align}
where $\kappa\simeq 7.6695$ and $I_k$, $K_k$ represent the modified Bessel functions. In particular, $\mathscr{B}(x)$ is very well approximated numerically by $\mathscr{B}(x)\simeq \sqrt{2\pi/x}(1/x+1)^{-2/3}$. A least squares fitting procedure \cite{Sofue2012} yields the parameter values $M_b=1.652\times 10^{10}M_\odot$, $a_b=0.522$~kpc, $M_d=3.41\times 10^{10}M_\odot$, $a_d=3.19$~kpc, $\rho_0=1.06\times 10^7M_\odot/$kpc, $h=12.53$~kpc.

\begin{figure}[t]
\begin{center}
\begin{minipage}{\linewidth}
\includegraphics[width=\linewidth]{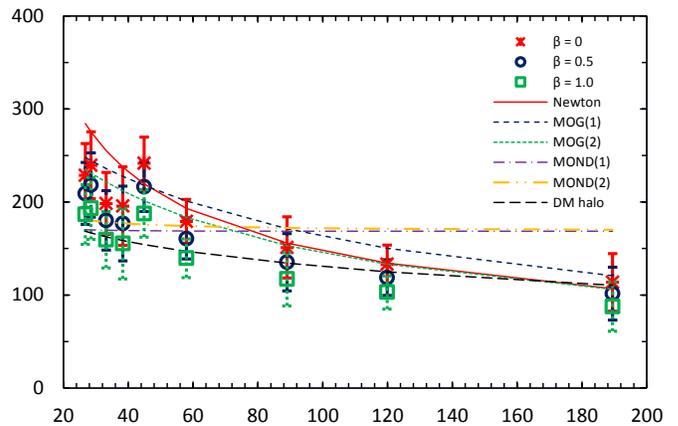}
\caption{Fits to the observed rotational velocities of the Milky Way galaxy using various theories. The data \cite{Bhattacharjee2014} are represented with red crosses ($\beta=0$), blue hollow circles ($\beta=0.5$) and green hollow squares ($\beta=1.0$). The solid red line is the Newtonian fit with a central mass of $M=5\times 10^{11}M_\odot$. The blue medium dashed and green short dashed lines correspond to MOG using the values of $M=4\times 10^{10}M_\odot$, $\alpha=15.01$, $\mu=0.0313$~kpc$^{-1}$, and $M=5\times 10^{10}M_\odot$, $\alpha=8.89$, $\mu=0.04$~kpc$^{-1}$, respectively, which correspond to the values calculated according to Ref. \cite{Moffat2007e}, or given by Ref. \cite{MoffatRahvar2013}. The purple dash-dotted line and the yellow dash-double-dot line correspond to fits using MOND, the mass of $M=5\times 10^{10}M_\odot$, $a_0=1.21\times 10^{-8}$~cm/s$^2$, and the interpolation functions given by Eqs.~(\ref{MOND1}) and (\ref{MOND2}), respectively. Finally, the black long-dashed line is the dark matter halo prediction given in \cite{Sofue2012}.\label{fig1}}
\end{minipage}
\end{center}
\end{figure}

The results, along with the data reported in Ref.~\cite{Bhattacharjee2014} are presented in Fig. \ref{fig1} and discussed below.

\section{Results and Discussion}

Figure~\ref{fig1} contains data from \cite{Bhattacharjee2014} and predictions of MOG, MOND, the dark matter halo and a Newtonian point source with an appropriately chosen mass.

The data are rotational velocities from a heterogeneous set of objects, as reported in Fig.~5 (lower right) in Ref.~\cite{Bhattacharjee2014}. As these are non-disk objects, the rotational velocities are adjusted by a factor of $1/\sqrt{2}$; this is discussed in \cite{Sofue2012}. Raw observations are converted into estimates of rotational velocities using three different values of the anisotropy parameter $\beta$.

As can be seen in Fig.~\ref{fig1}, the MOG prediction is in agreement with the data using either set of the parameters $\alpha$ and $\mu$. The source mass was chosen correspondingly: the mass-dependent values of $\alpha=15.01$, $\mu=0.0313$~kpc$^{-1}$ were obtained for a source mass of $M=4\times 10^{10}M_\odot$, whereas the ``universal'' values of $\alpha=8.89$ and $\mu=0.04\,{\rm kpc}^{-1}$ offered a good fit with $M=5\times 10^{10}M_\odot$.

The Newtonian point source rotation curve, with a source mass of $M=5\times 10^{11}M_\odot$, clearly corresponds to an unrealistic $\Upsilon$ in the absence of dark matter. Nonetheless, the shape of this rotation curve calls attention to the fact that the actual rotation curve is {\em not flat}. The Milky Way rotation curve {\em does} follow the Keplerian prediction at large radii, being proportional to the inverse square root of the radial distance.

In contrast, MOND predicts a flat rotation curve, just as it is designed to do, and regardless of the choice of the interpolation function. The only way to force MOND to yield a Keplerian rotation curve would be by reducing the value of the MOND acceleration parameter $a_0$ to such an extent that the MOND prediction becomes indistinguishable from the Newtonian prediction, and requires an equally unrealistic value of $\Upsilon$. One possible way to reconcile the tension between the MOND prediction and the data might be the introduction of a variable anisotropy parameter $\beta(r)$, dependent on radial distance; however, as can be discerned by comparing the data points that correspond to the assumption of $\beta=0$ vs. $\beta=1$ in Fig.~\ref{fig1}, even a drastically varying $\beta(r)$ would offer at best marginal agreement between MOND and the data set.

As anticipated, the dark matter halo profile fits the data well; this is hardly surprising, given that this profile has as many as six independently fitted parameters characterizing the bulge, the disk and the halo.

\section{Conclusions}

Extended rotation curve observations of the Milky Way galaxy up to a radial distance of 200~kpc show Keplerian behavior, with rotational velocities being proportional to the inverse square root of the radial distance. These data are strongly incompatible with the predictions of the MOND paradigm, as they show a Keplerian dependence of the rotational velocity on radial distance. In contrast, Scalar-Tensor-Vector Gravity (STVG), also known as modified gravitational (MOG) theory, can be used to fit the new data easily, with no changes to the previously established parameters of the theory; in contrast, the extended dark matter halo is a fitted model with as many as six parameters that are independently adjusted to match observations.

\section*{Acknowledgments}

This research was generously supported by the John Templeton Foundation. Research at the Perimeter Institute for Theoretical Physics is supported by the Government of Canada through industry Canada and by the Province of Ontario through the Ministry of Research and Innovation (MRI).

\bibliography{refs}

\end{document}